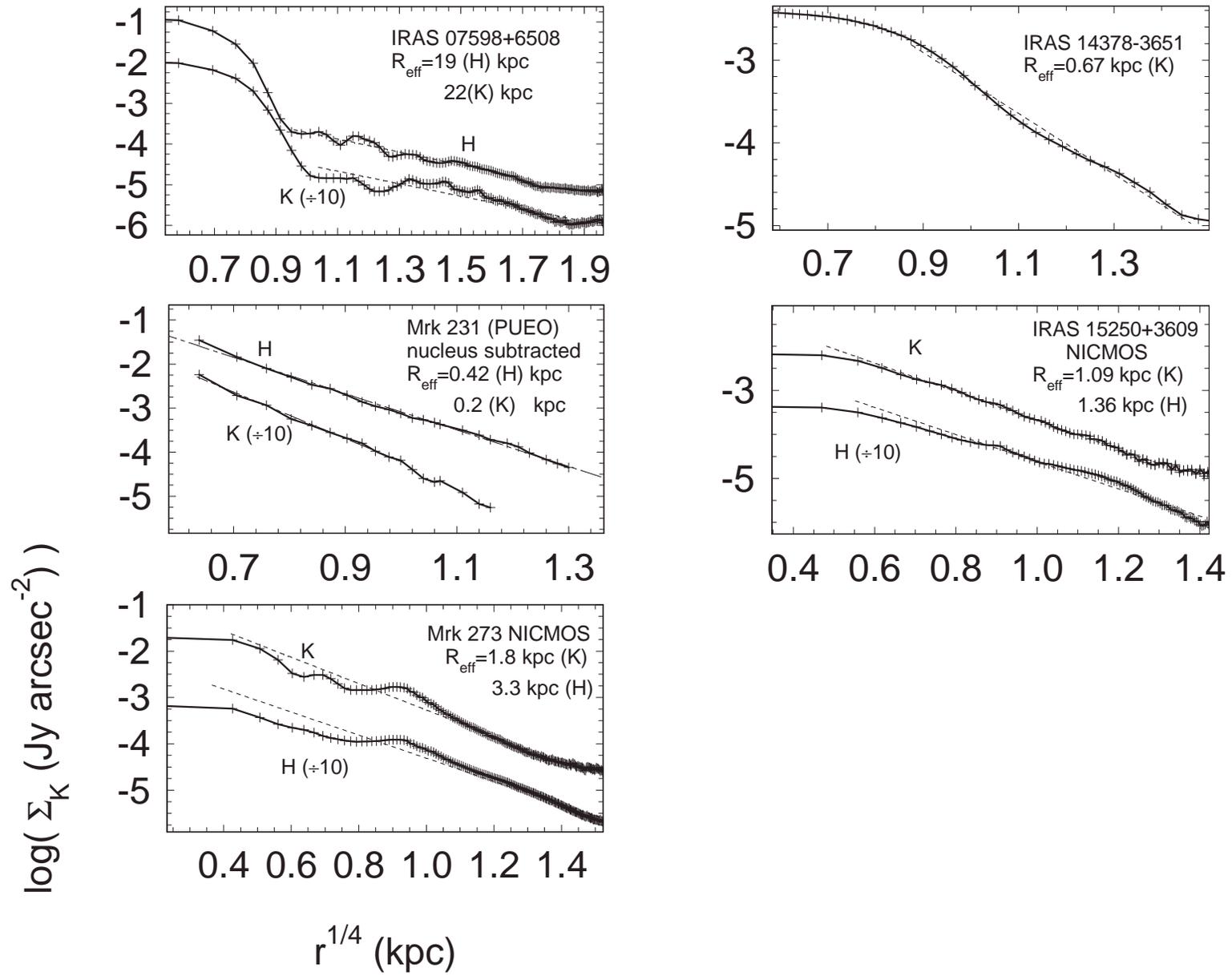

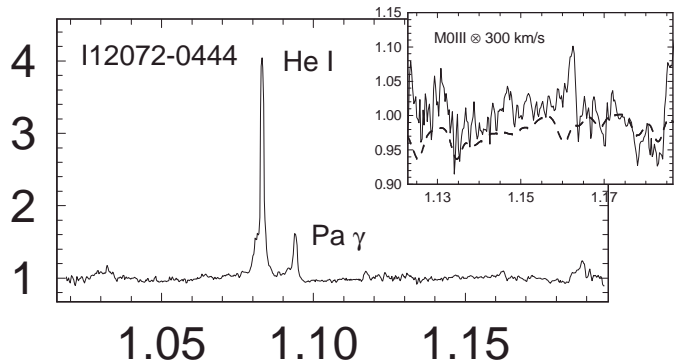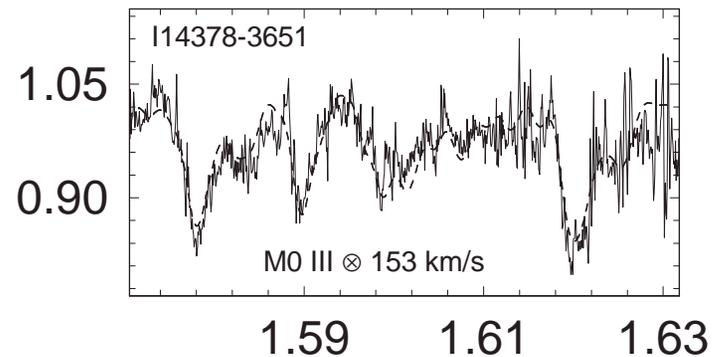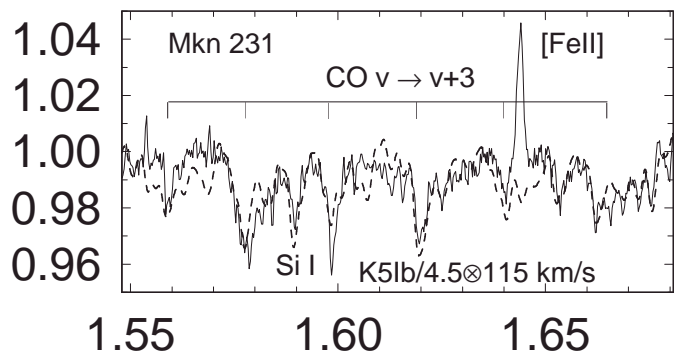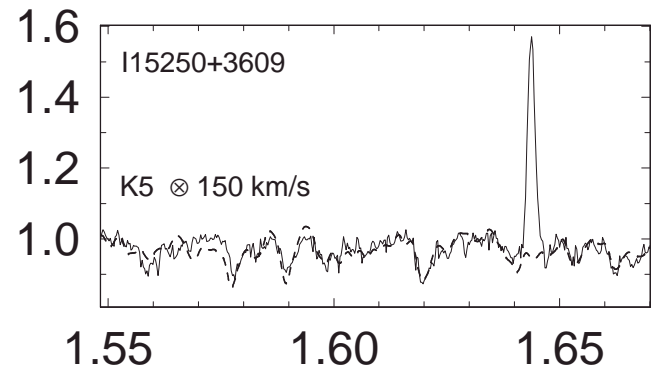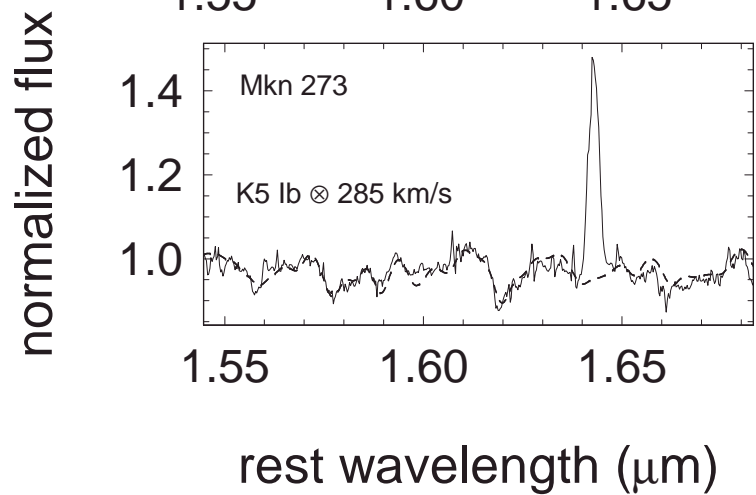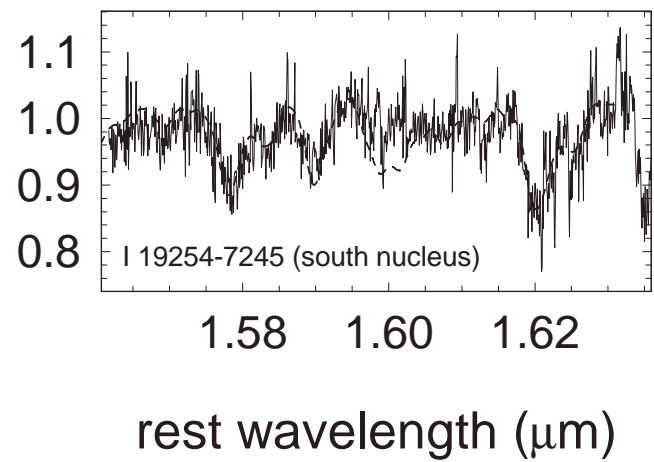

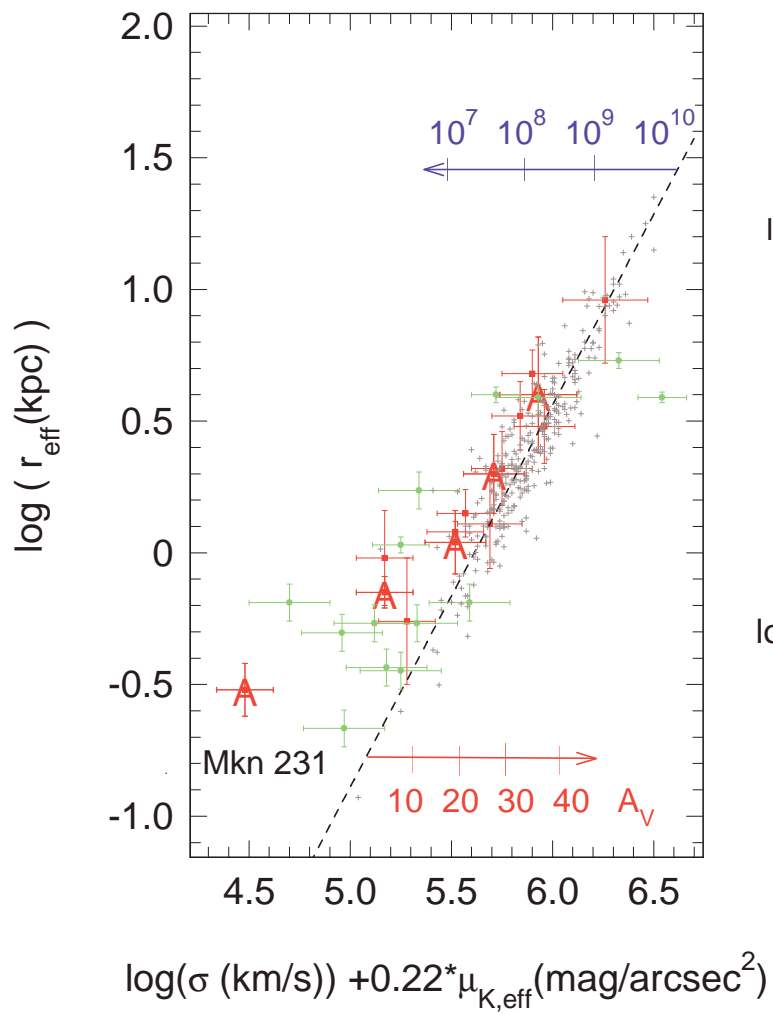
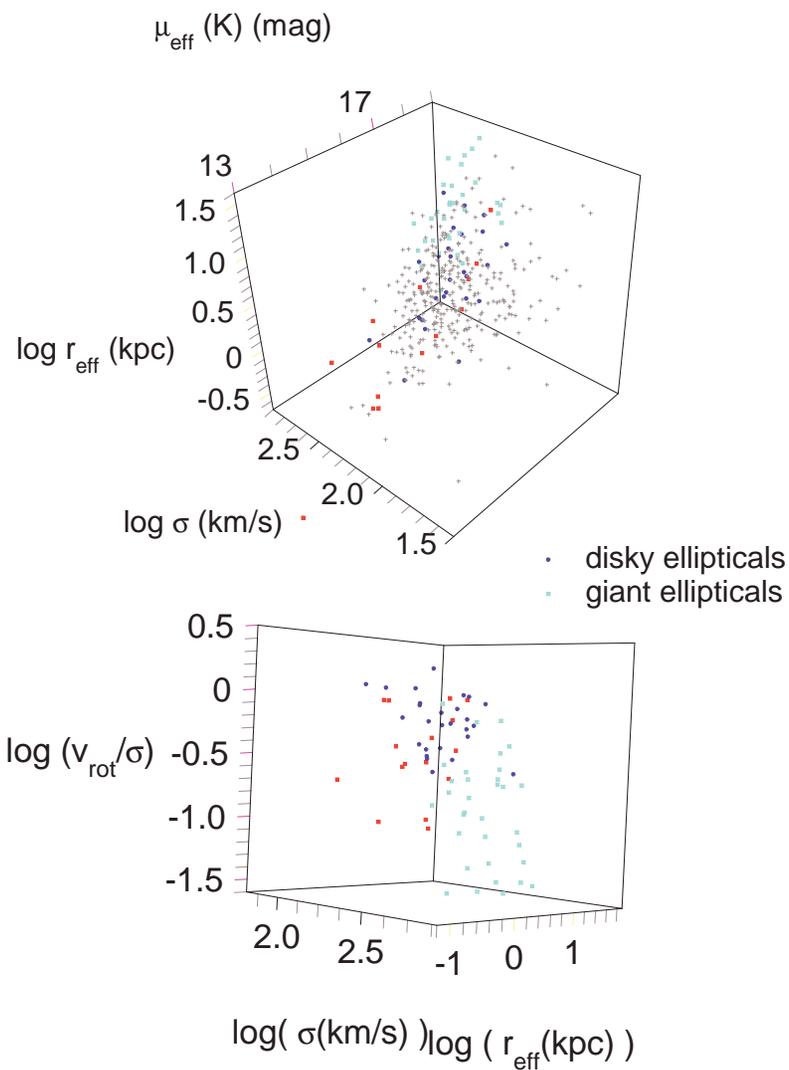

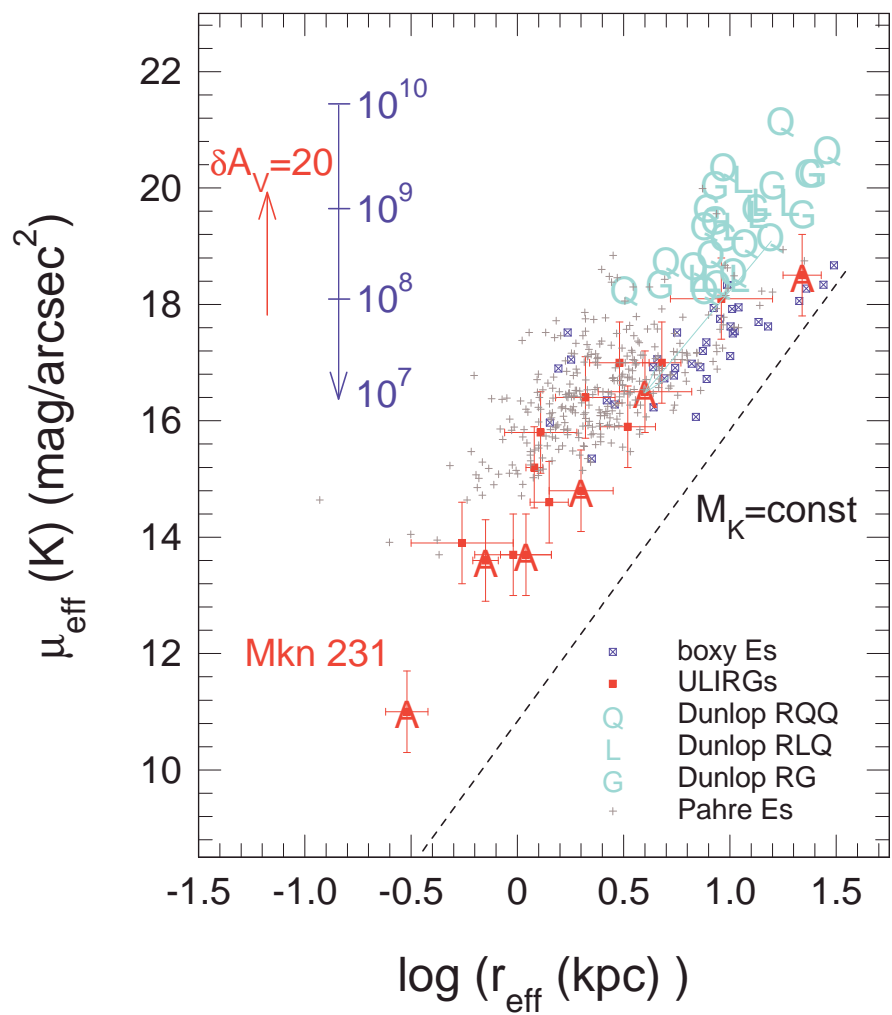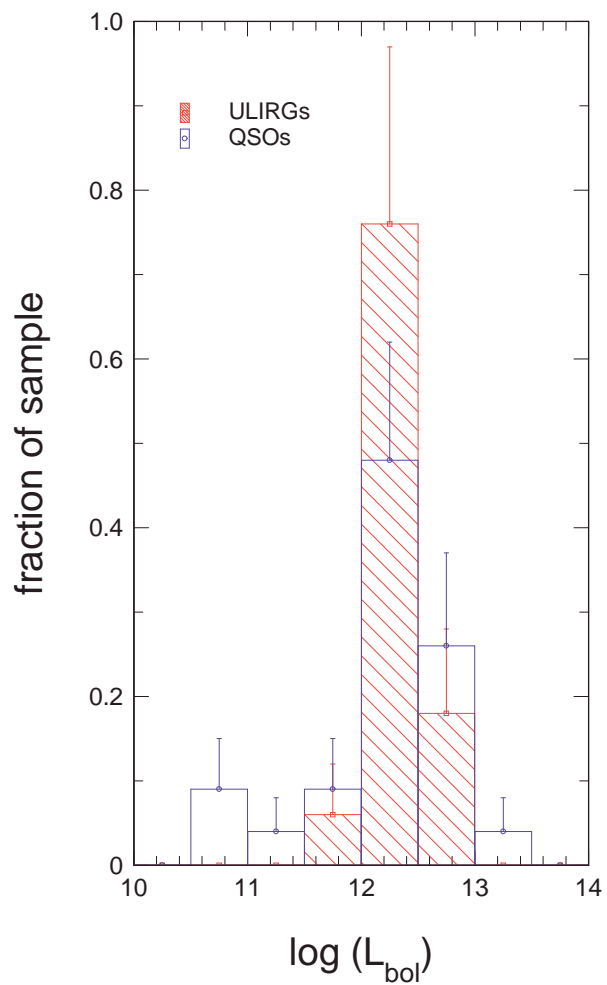

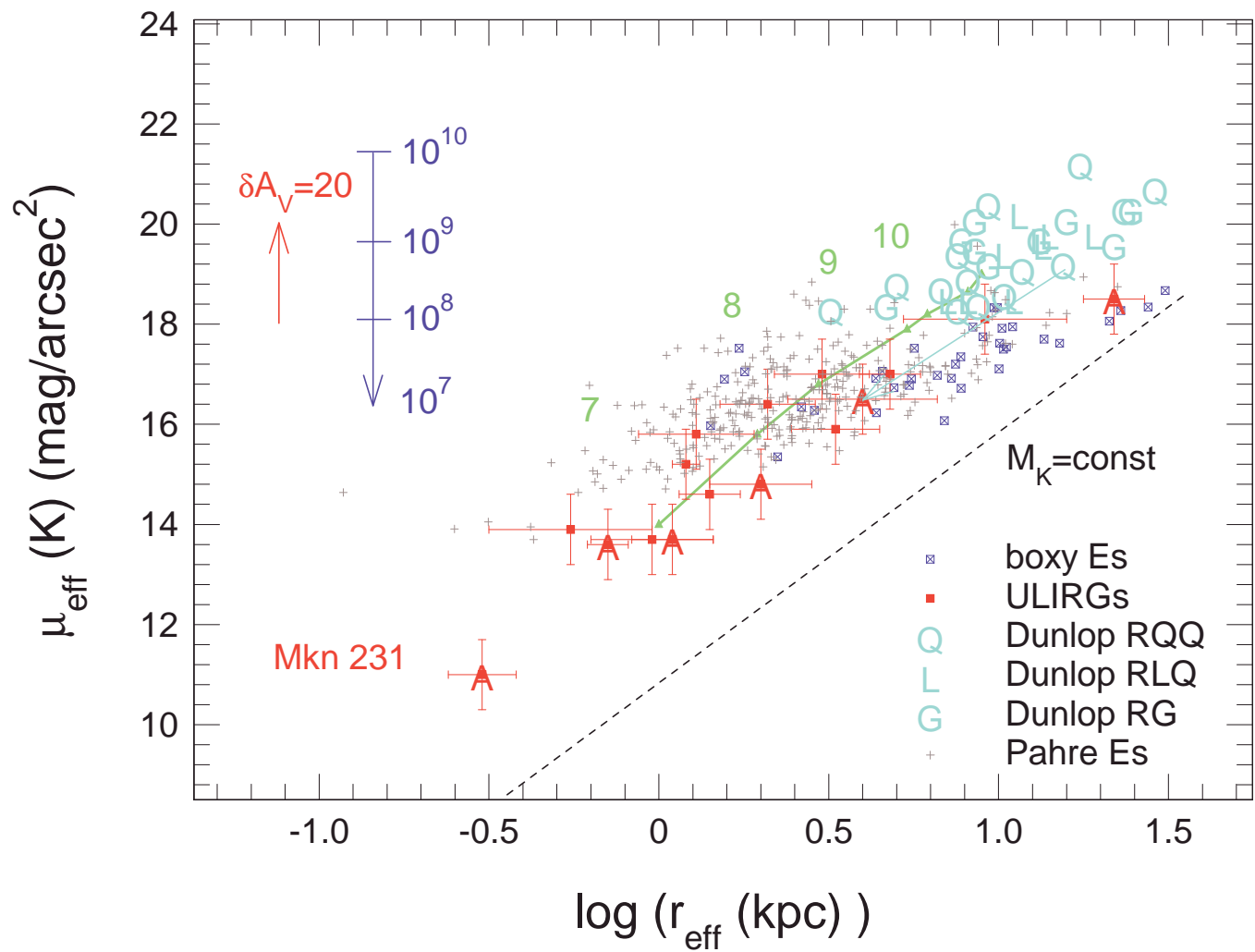

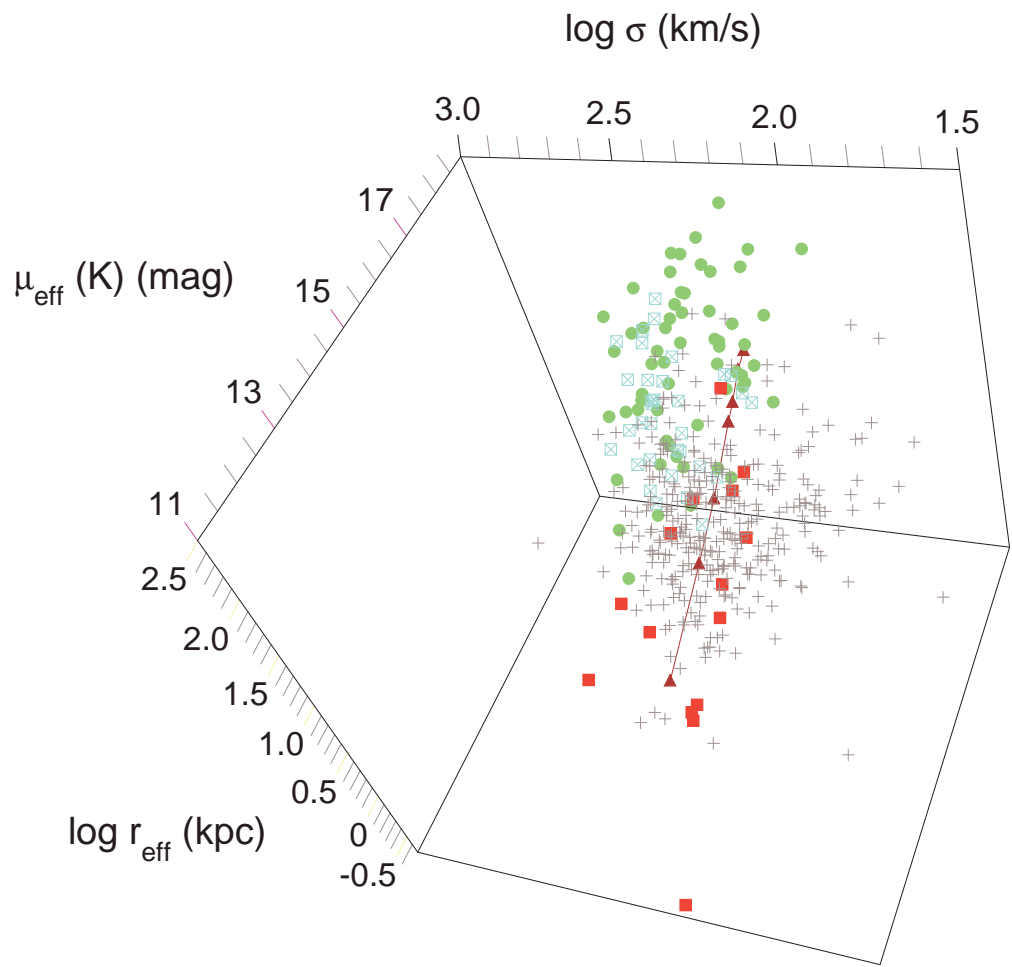

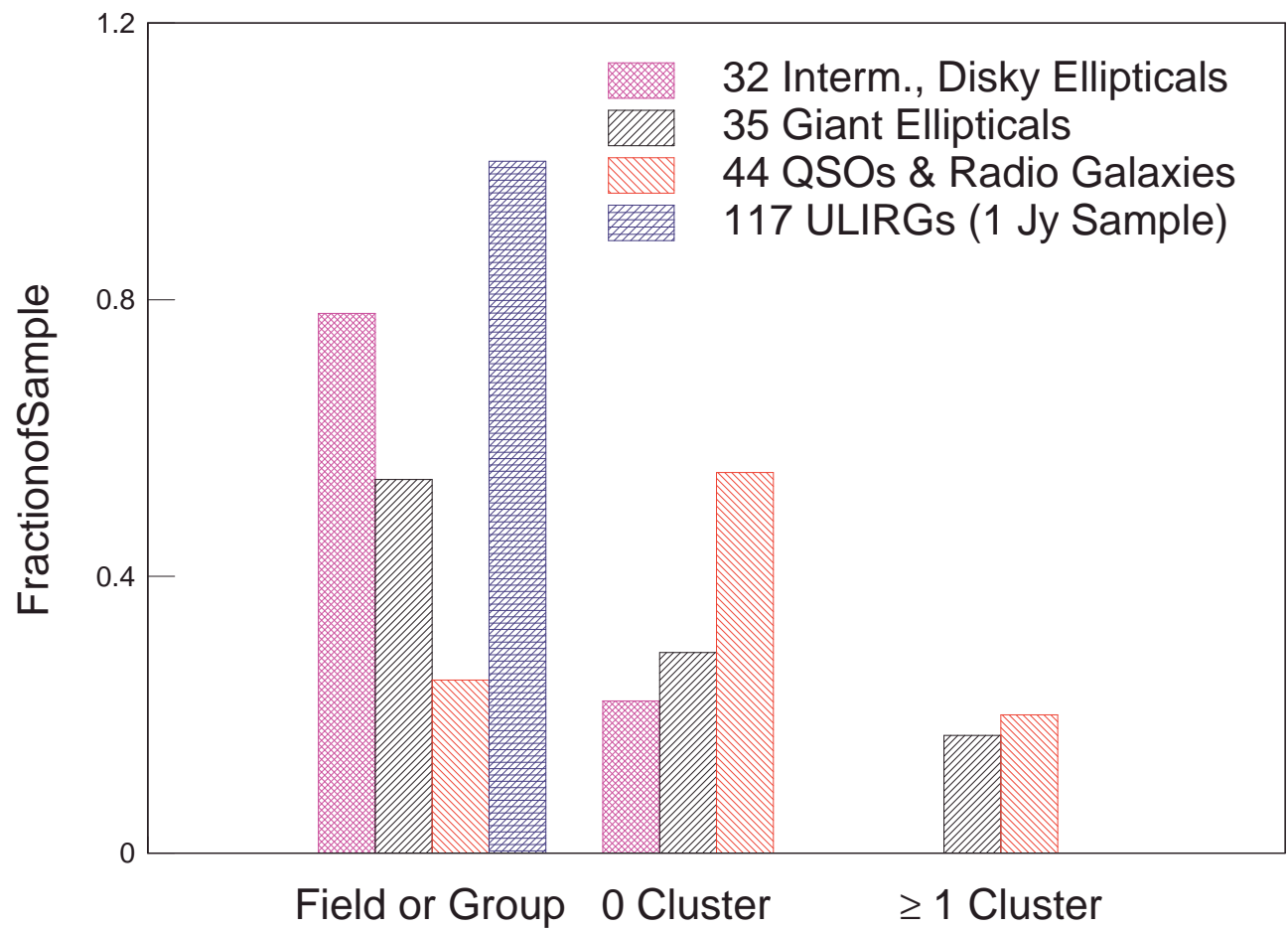

# Ultra-Luminous Infrared Galaxies: QSOs in Formation?[1]

L.J.Tacconi[1], R.Genzel[1,2], D.Lutz[1], D.Rigopoulou[1], A.J. Baker[1],

C.Iserlohe[1] and M.Tecza[1]

[1]Max-Planck Institut für extraterrestrische Physik, Garching, FRG
[2]Dept. of Physics, University of California, Berkeley, USA

## ABSTRACT

We present new near-infrared Keck and VLT spectroscopic data on the stellar dynamics in late stage, ultra-luminous infrared galaxy (ULIRG) mergers . We now have information on the structural and kinematic properties of 18 ULIRGs, 8 of which contain QSO-like active galactic nuclei. The host properties ($\sigma$, $r_{eff}$, $\mu_{eff}$, $M_K$) of AGN-dominated and star formation dominated ULIRGs are similar. ULIRGs fall remarkably close to the fundamental plane of early type galaxies. They populate a wide range of the plane, are on average similar to $L_*$-rotating ellipticals, but are well offset from giant ellipticals and optically/UV bright, low-z QSOs/radio galaxies. ULIRGs and local QSOs/radio galaxies are very similar in their distributions of bolometric and extinction corrected near-IR luminosities, but ULIRGs have smaller effective radii and velocity dispersions than the local QSO/radio galaxy population. Hence, their host masses and inferred black hole masses are correspondingly smaller. The latter are more akin to those of local Seyfert galaxies. ULIRGs thus resemble local QSOs in their near-IR and bolometric luminosities because they are (much more) efficiently forming stars and/or feeding their black holes, and not because they have QSO-like, very massive black holes. We conclude that ULIRGs as a class cannot evolve into optically bright QSOs. They will more likely become quiescent, moderate mass field ellipticals or, when active, might resemble the X-ray bright, early type galaxies that have recently been found by the Chandra Observatory.

---

[1] Based on observations at the European Southern Observatory, Chile, and on observations at the W.M. Keck Observatory, which is operated as a scientific partnership among the California Institute of Technology, the University of California and the National Aeronautics and Space Administration. The Keck Observatory was made possible by the general financial support by the W.M. Keck Foundation.

*Subject headings: galaxies: elliptical and lenticular, cD – galaxies: active – galaxies: kinematics and dynamics – infrared: galaxies- quasars: general*

# 1. INTRODUCTION

The X-ray and far-IR/submm extragalactic backgrounds show that dust-enshrouded starbursts and AGN play an important role in the high redshift Universe (Pei, Fall, & Hauser 1999; Miyaji, Hasinger & Schmidt 2000). The extragalactic far-IR/submillimeter background (Hauser & Dwek 2001, and references therein) appears to be dominated by luminous (>$10^{11.5} L_\odot$) galaxies at $z \geq 1$ (e.g., Smail, Ivison & Blain 1997; Hughes et al. 1998, Barger et al. 1998; Genzel & Cesarsky 2000; Elbaz et al. 2002). Very luminous and massive (Rigopoulou et al. 2002), dusty starbursts (star formation rates $10^2$ to $10^3$ $M_\odot yr^{-1}$) thus have probably been contributing significantly to the cosmic star formation rate at $z \geq 1$. These dusty starbursts may be large bulges/ellipticals in formation (e.g., Guiderdoni et al 1998; Blain et al. 1999).

(Ultra-) luminous infrared galaxies ((U) LIRGs[2]: Sanders et al 1988, Sanders & Mirabel 1996) may be the local analogues of the high-z population. Following the 'ellipticals through mergers' scenario of Toomre & Toomre (1972) and Toomre (1977), Kormendy & Sanders (1992) proposed that ULIRGs might evolve into ellipticals through merger-induced, dissipative collapse. Invariably ULIRGs are advanced mergers of gas-rich disk galaxies (Sanders & Mirabel 1996). In the process of merging such systems may go through a luminous starburst phase and later evolve into QSOs (Sanders et al. 1988).

---

[2] In the definitions of Soifer, Houck & Neugebauer (1987) and Sanders & Mirabel (1996) 'ultra-luminous' infrared galaxies (ULIRGs) have infrared (10-1000μm) luminosities >$10^{12} L_\odot$. Galaxies with infrared luminosities <$10^{12}$ and >$10^{11}$ $L_\odot$ are called 'luminous' (LIRGs). While these definitions are somewhat arbitrary, only the ULIRG category is comprised almost exclusively of compact mergers of very gas rich and luminous galaxies.

ULIRGs often contain active galactic nuclei[3], and some fraction of these systems attains QSO-like bolometric luminosities (Sanders et al. 1988; Genzel et al. 1998; Lutz, Veilleux & Genzel 1999; Veilleux, Kim & Sanders 1999). Formation of ellipticals through major mergers and triggering of major starbursts and QSOs may thus be intimately coupled. ULIRGs have large central molecular gas concentrations with densities comparable to stellar densities in ellipticals (Downes & Solomon 1998; Bryant & Scoville 1999; Sakamoto et al. 1999; Tacconi et al. 1999). These gas concentrations (and stars forming from them), along with dense and massive central bulges in the precursor galaxies (Hernquist, Spergel & Heyl 1993), may be the crucial ingredients for overcoming the fundamental phase space density constraints that would otherwise prevent formation of dense ellipticals from much lower density disk galaxies (Ostriker 1980).

In a previous paper (Genzel et al. 2001), we have begun to test the 'elliptical-in-formation'-scenario by determining the fundamental structural and kinematic properties of the stellar hosts of late stage ULIRG mergers: velocity dispersion and rotation, effective (~ half light) radius, mean surface brightness, and absolute host magnitude. If ULIRGs evolve into ellipticals they should lie on the fundamental plane of early type galaxies (Djorgovski & Davis 1987; Dressler et al. 1987). While ULIRGs and mergers as a class are transitory objects undergoing rapid evolution, late stage mergers probably provide a fair sample of the dynamical properties of the systems into which they will finally evolve. Recent numerical simulations of galaxy mergers suggest that violent relaxation is very effective. Once two nuclei merge to a separation of $\leq 1$ kpc, the stellar system has basically reached its equilibrium values of rotation, dispersion and higher order kinematic moments (skewness and kurtosis) on spatial scales of the half mass radius or greater, and coalescence is rapid (Mihos 1999; Bendo & Barnes 2000). Thus, although the structural parameters are necessary to place ULIRGs on the Fundamental Plane of early-type galaxies, it is the quantitative comparisons of the dynamical parameters that will be the most robust. Since ULIRGs are dusty and their central few kpc regions are usually highly obscured ($A_V \geq 5$), their dynamical properties must be measured in the near-infrared. Most are moderately distant (z~0.1), and high quality spectra (SNR on the continuum of 50 to 100) are necessary to extract reliable velocity dispersions. Such observations require 10m-class telescopes and sensitive near-IR spectrometers.

---

[3] Active galactic nuclei are here defined to be objects whose bolometric luminosity is derived from accretion onto a massive central black hole.

In Genzel et al. (2001) we have reported high quality stellar (and gas) near-IR spectroscopy of a modest sample of late-stage (single nuclei, or compact double nuclei) ULIRGs , taken with the Keck telescope and the VLT. These data strongly support the 'elliptical-in-formation' scenario. ULIRG mergers indeed fall near the fundamental plane and resemble moderate mass , rotating (disky), L* ellipticals and lenticular galaxies. In this second paper, we present further spectroscopy of ULIRGs and test the second part of the Sanders et al. (1988) scenario: Once rid of their gas and dust shells, are ULIRGs the likely progenitors of optically/UV bright, 'naked' QSOs?

## 2. OBSERVATIONS AND DATA REDUCTION

The data were taken with the Keck 2 telescope on Mauna Kea, Hawaii, and with the ANTU telescope of ESO's VLT on Cerro Paranal, Chile. At the Keck 2 telescope, the data were taken with the facility spectrometer NIRSPEC (McLean et al. 1998) in spring 2001. At the VLT the data were taken with the facility near-infrared camera and spectrometer ISAAC (Moorwood et al. 1998) in two observing runs in spring and summer 2001. NIRSPEC was used in low-resolution mode and had R(FWHM)=$\lambda/\Delta\lambda$=2200, a slit width of 0.58", and 0.193" per pixel along the slit. ISAAC was used in medium-resolution mode (R (FWHM)=5200), with a slit width of 0.6" and 0.147" per pixel along the slit.

As in Genzel et al. (2001), we selected galaxies fitting the ULIRG criteria ($L_{IR} \geq 10^{12} L_\odot$) from the BGS (Sanders et al. 1988; Sanders & Mirabel 1996), 2 Jy (Strauss et al. 1992), and 1 Jy (Kim & Sanders 1998) catalogues. We culled from these IR-luminosity selected catalogues those sources fitting our RA/Dec boundary conditions, having single nuclei or compact (projected separation < a few kpc) double nuclei in near-IR images, and having definite signatures of a recent merger like tidal tails. Such systems are very likely late stage merger remnants. We picked only those sources with redshifts ≤0.16, for which reasonably strong stellar absorption features fall in the H- or J-bands (see Genzel et al. 2001). These constraints reduced the number of available sources from about 70 (fitting the RA/Dec-requirements) to about 20. To address the specific issue of AGN-dominated ULIRGs, we concentrate in this paper on sources known to have strong or energetically dominant AGN from previous infrared and optical spectroscopy or spectrophotometry (Sanders et al. 1988, 1989; Duc, Mirabel & Maza 1997). We were able to get spectra of seven sources. These ULIRG mergers are all located in the field or in small groups; none are situated in dense clusters (see section 4.3). Table 1 gives the source list. All distances in this paper are for a $H_o$=70 km/s/Mpc, $\Omega_m$=0.3, $\Omega_{tot}$=1 cosmology.

For the data reduction and extraction of stellar spectra and kinematics, we used the same techniques as in our previous work. We refer to Genzel et al. (2001) for the full description. In the case of the NIRSPEC data, the flat fields and some of the data (taken with a new InSb detector in use since December 2000) suffered from saturation effects. For this reason, we did not divide by the flat field (the new detector is of excellent cosmetic quality and homogeneous response) and corrected the affected data sets by a saturation curve kindly provided to us by David Sprayberry.

## 3. RESULTS

### 3.1 Structural and dynamical properties of the ULIRG sample

The basic results of our study are contained in Figures 1 and 2 and summarized in Table 2. Figure 1 shows the H- and K-band surface brightness distributions as a function of $r_{major-axis}^{1/4}$ for five of the galaxies kindly provided to us by N. Scoville (Scoville et al. 2000) and H. Dannerbauer (priv.comm.), and taken from Lai et al. (1998), along with the derived effective radii of the best de Vaucouleurs ($r^{1/4}$) law fits. Figure 2 displays our final spectra, and comparisons with K/M stellar template spectra convolved with the Gaussians best reproducing the line of sight velocity distributions. Table 2 summarizes for our sources the derived stellar velocity dispersions, rotation velocities for stars and gas, effective radii, average K-band surface brightness within $r_{eff}$, $\mu_{eff}$, and integrated, observed absolute K-band magnitudes. Appendix A contains comments on individual sources.

As was true for the sample of Genzel et al. (2001), most of the merger remnants in Table 2 have compact near-infrared light distributions, with effective radii of a few kpc or less. The effective radii and surface brightnesses listed in Table 2 have substantial uncertainties (±30% to ±50%), because of differential extinction, incomplete relaxation and population effects. NIR images show significant deviations from ellipsoidal shapes in many sources (including the tidal tails: Duc et al. 1997; Rigopoulou et al. 1999; Scoville et al. 2000; Surace, Sanders & Evans 2000; Veilleux, Kim & Sanders in preparation; Bushouse et al. 2001). Deep K-band CO absorption features in these sources indicate that relatively young stars are present and may affect the surface brightness distributions (10 to a few 100 Myr: Armus et al. 1995; Goldader et al. 1995; Tecza et al. 2000). The greatest source of uncertainty, however, is extinction. Equivalent screen, K-band extinctions range between 0.4 and >2 mag, as determined from near-infrared colors or infrared line ratios (Scoville et al. 2000; Genzel et al. 1998; section 4.4). Colors redden towards the nuclear regions in most of our ULIRGs (Scoville et al. 2000; Fig.1). K-band derived effective radii are thus generally smaller than H-band effective radii, and these in turn are smaller than I-/R-/V-band effective radii (Fig.1). The most likely explanation of this finding is that extinction increases at smaller radii. Other possibilities may be nuclear hot dust emission (e.g. from the AGN), or a circum-nuclear concentration of young, cool supergiants. In Table 2, we again used averages of the K- and H-band data (if available) to determine effective radii, but also excluded the nuclear regions (<400 pc) from our fits (see Fig.1), in order to avoid any biases due to compact central sources associated with the AGN in our sample.

The fits in Figure 2 of the Gaussian-convolved K5/M0 template star spectra to the galaxy spectra are generally very good, even in the case of Mkn 231, whose AGN hot dust continuum dilutes the stellar continuum by a factor of 4.5 and reduces absorption feature depths to less than 4%. One exception is the source IRAS 12071-0444, for which strong (AGN) emission lines dominate the observed J-band spectrum, and the stellar absorption features are weak and poorly constrained. In the case of IRAS 19254-7245 (the 'Superantennae', e.g. Sanders & Mirabel 1996), we were only able to measure the velocity dispersion of the southern nucleus, and therefore do not consider this source further in the discussion below. We have been able to derive gas rotational velocities either from NIR emission lines (this paper and Genzel et al. 2001) or from CO millimeter interferometry from the literature (Downes et al 1998; Sakamoto et al. 1999; Tacconi et al. 1999) for about half of our total ULIRG sample. Downes et al (1998) have measured millimeter CO velocity dispersions for 5 objects in our sample.

### 3.2 Location of the ULIRG mergers relative to the fundamental plane

We list the structural and dynamical parameters of the ULIRGs both from this work and from Genzel et al. (2001) in Table 3. Figure 3 shows the location of the full set of ULIRG mergers in two different projections of the fundamental plane of early type galaxies, as spanned by the samples of Pahre (1999), Bender, Burstein & Faber (1992), and Faber et al. (1997). We mark ULIRGs with strong or energetically dominant AGN with an "A" in the figure. We also show the location of the ULIRGs in a plane spanned by $r_{eff}$, $\sigma$ and $v_{rot}/\sigma$, as compared to slowly rotating (~boxy) ellipticals and moderately fast rotating (~disky) ellipticals and S0s/lenticulars.

The new sources improve our statistics and further strengthen the conclusion of Genzel et al. (2001) that late-stage ULIRG mergers are located very close to or in the fundamental plane. Their properties are similar to those of moderate mass (~$L_*$) ellipticals/lenticulars with moderately fast rotation, but differ significantly from those of giant ellipticals. The latter have significantly larger velocity dispersions and effective radii than the ULIRGs. These new results further suggest that AGN-dominated ULIRGs have dynamical properties similar to those of ULIRGs powered predominantly by star formation. Both types of ULIRG have stellar dynamics that is relaxed, but where rotation plays a significant role. The gas and stellar motions are often decoupled, as found by Genzel et al. 2001. While the gas rotational velocities are often greater than those of the stars, the CO velocity dispersions tend to be lower than those of the stars. The dynamical decoupling of the stars and gas is expected, since the stars have likely undergone strong dynamical heating due to violent relaxation, while the gas dissipates to a cold state on a dynamical timescale. Beam smearing effects may play a greater role in the gas velocities derived from the CO data, since these have lower spatial resolution than the Keck and VLT NIR spectra. We will present a detailed comparison of the stellar and gas dynamical properties derived from new millimeter interferometry CO measurements of our northern ULIRG sample in a forthcoming paper (Baker et al., in preparation).

Taken together with somewhat lower luminosity (LIRG) mergers (green symbols in Figure 3; from Shier & Fischer 1998; James et al. 1999), these sources span wide ranges in both dimensions of the fundamental plane. One obvious outlier is Mkn 231. This QSO/ULIRG has both a very low velocity dispersion (with little observed rotation) and a very small host effective radius, yet a large near-infrared surface brightness. We discuss this source in more detail in Appendix A.

The proximity of the ULIRG mergers to the fundamental plane is somewhat surprising (Genzel et al. 2001). The merger remnants are known to have undergone active star formation for the past 10 to a few $10^2$ Myr (Goldader et al. 1995, 1997; Armus et al. 1995; Genzel et al. 1998; Tecza et al. 2000). Their near-IR emission is likely affected and in a few cases even dominated by red supergiants (Armus et al. 1995; Tecza et al. 2000). As shown by the blue arrow in the left inset of Figure 3, the presence of such a young stellar population ought to significantly enhance the surface brightness relative to that of a $>10^{10}$ Gyr population (by 5 magnitudes in the supergiant phase, and 2 to 3 magnitudes in the AGB phase: Bruzual & Charlot 1993). The most obvious interpretation of the close proximity of the ULIRGs to the fundamental plane is thus a 'cosmic conspiracy' of stellar evolution and extinction. To 'undo' the expected surface brightening of a few hundred Myr old population relative to the old population in ellipticals, a V-band screen extinction of 20 to 30 magnitudes is required. Such values of obscuration are plausible given the near-IR colors and mid-IR emission line ratios of ULIRGs (e.g. Genzel et al. 1998; and section 4.4). This interpretation is also consistent with the fact (Figure 3, left inset) that the sources with the smallest effective radii, likely the youngest ULIRG and LIRG mergers, are also the most deviant from the fundamental plane. Although extinction and population effects likely play a major role in determining the photometric and structural quantities of ULIRGs the dynamical similarities of ULIRGs and intermediate mass elliptical galaxies are robust, since the late stage ULIRGs should already be close to their final dynamical state (e.g. Mihos 1999).

## 4. DISCUSSION: DO ULIRG MERGERS EVOLVE INTO QSOs?

### 4.1 Comparison of ULIRGs to optically bright QSOs/radio galaxies

In the scenario of Sanders et al. (1988), a dusty ULIRG merger is initially powered by star formation, but increasingly has its bolometric luminosity dominated by accretion onto the central black hole(s) as the merger progresses. As the dust and gas are cleared out in the last phases of the merger, the dusty (infrared excess) QSO evolves into a classical, optically bright QSO.

If this scenario is correct, late stage ULIRGs (likely AGN dominated) and optically bright QSOs of similar luminosities should have similar host properties. To test whether this is the case, we compare our ULIRG sample to the sample of 33 radio-quiet and radio-loud QSOs and radio galaxies (z<0.25) studied in detail with the HST and ground-based telescopes by Dunlop et al. (2002, and references therein). The right inset in Figure 4 indeed confirms that the nuclear (bolometric) luminosities of the radio-quiet and radio-loud QSOs of the Dunlop et al. sample are basically identical to those of the total ULIRG sample, and thus that a direct comparison of their hosts is appropriate (see also e.g. Soifer et al 1986, Soifer, Houck & Neugebauer 1987) for those cases where AGN make a substantial contribution to the bolometric luminosity of the ULIRGs. For this comparison, we have taken the nuclear absolute R-band magnitudes of Dunlop et al., converted these to our cosmology, and applied a bolometric correction of a factor of 14 (2.9 mag), as obtained from the average

QSO spectral energy distribution of Elvis et al. (1994). In the case of the ULIRGs, we computed the 1-1000μm luminosity from the IRAS fluxes using the recipe of Sanders & Mirabel (1996) and applied a bolometric correction of 25% to account for the near-IR and optical contributions to the luminosity. This results in a lower limit to the bolometric luminosities of the ULIRGs, since no allowance is made for escaping radiation of extreme UV radiation. On the other hand, in ≥75% of the ULIRGs star formation accounts for ≥50% of the bolometric luminosity, thus reducing the AGN contribution to the bolometric luminosity accordingly (Genzel et al. 1998). In optically bright QSOs the bolometric correction for the mid- to far-IR luminosity is a factor of 5 (Elvis et al. 1994).

For their optically selected AGN Dunlop et al. have determined effective radii from HST R-band imaging, and K-band average surface brightness values (within $r_{eff}$) from the HST data and ground-based/HST R-K values. There are presently no dynamical measurements for these AGN, so we can compare our ULIRG and the Dunlop et al. QSO/radio galaxy data only in the $\mu_{eff}$-$r_{eff}$ projection of the fundamental plane (Figure 4). As emphasized by Dunlop et al., QSOs and radio galaxies are located in the upper right of the diagram, coincident with the locus of giant ellipticals, and with typical effective radii of ≥10 kpc. Bahcall et al. (1997) have measured V-band effective radii with HST for a sample of 20 z<0.3 QSOs, and find a mean effective radius of 7.5 kpc converting to our cosmology. In contrast ULIRGs have effective radii of a few kpc or less, and much higher surface brightness. ULIRGs and the Dunlop et al. QSOs/radio galaxies thus are located in different parts of the fundamental plane.

## 4.2 Differential extinction and population effects

To be sure that this obvious difference is truly caused by different host properties we need to consider again the effects of extinction and evolution on the location of sources in the $\mu_{eff}$-$r_{eff}$ plane. In the presence of large visible extinction and hidden circum-nuclear star formation regions, effective radii may be wavelength dependent, as appears to be the case for many ULIRGs. In fact, one of the radio-quiet QSOs studied by Dunlop et al. is Mkn 1014, which is also in our sample. For this source the effective radius decreases from 15 kpc in the R-band, to 4±2 kpc in the H/K-band (Genzel et al. 2001). This decrease in effective radius results in a substantial shift of Mkn 1014 in Figure 4. Mkn1014 is unique, however, in that it is the only (dusty) infrared excess QSO in the Dunlop et al. sample. We believe that Mkn 1014 is atypical for QSOs, but it shows what can happen in the presence of substantial differential extinction. A significant, but much smaller (~20%) difference between V-band and K-band effective radii also exists for the ellipticals (Pahre et al. 1998). In those sources, however, the difference is likely caused by a metallicity gradient rather than differential dust extinction.

Millimeter observations of CO rotational transitions have demonstrated that the molecular gas in ULIRGs is concentrated in a region of a few kpc or less (Downes & Solomon 1998; Bryant & Scoville 1999; Sakamoto et al. 1999; Tacconi et al. 1999), presumably as a result of the rapid loss of angular momentum during the merger process (Barnes & Hernquist 1992). These dense circum-nuclear molecular gas concentrations are coincident with the bright, highly obscured emission from young stars observed in the near-IR (Scoville et al. 2000), which may be superposed on more extended stellar light from the precursor disks.

During the dust-enshrouded initial merger phases the stellar light may be dominated by the central young stellar population, resulting in a small effective radius. Later, the near-IR emission of the young population will fade and the more extended and massive older population may begin to dominate, leading to an increase in effective radius.

To test this hypothesis, we constructed a toy model with a composite stellar population, consisting of an extended, old stellar population typical of the dominant stellar population in intermediate mass early-type galaxies ($r_{eff}$=10 kpc; age >$10^{10}$ yr) and a compact young population typical of the extreme central starburst regions observed in ULIRGs ($r_{eff}$=0.4 kpc; age ~$2x10^7$ yr). Both populations were assumed to follow an $r^{1/4}$-law distribution due to rapid violent relaxation during the merger. We chose the initial brightness ratio of the young and old components so as to match the location of the most compact ULIRG remnants in Figure 4 ($\mu_{eff}(K)$~14, $r_{eff}$~1kpc). We then let the K-band light of the young component evolve according to the Bruzual & Charlot (1993) models, empirically determining the effective radius of the merger remnant at each step. The resulting evolutionary track of the toy model in the $\mu_{eff}$-$r_{eff}$ plane as a function of age is shown in Figure 5. In this simple model we have not accounted for any additional star formation from the yet unused reservoir of molecular gas. Taking gas into account will not change the location of the toy model evolution vector, however. The effect would rather be to slow down or delay the evolution along the track until all of the gas is either consumed or expelled. The toy model does an excellent job in connecting the compact ULIRG remnants to the much more extended QSOs/radio galaxies of the Dunlop et al. sample, suggesting that an increase of the effective radius by a factor of 10, toward the locus of giant ellipticals, is in principle possible during the evolution of the merger remnant.

We nevertheless think that it is unlikely that such an extreme evolutionary shift is typical for most ULIRG merger remnants in Figure 5. The toy model makes the extreme assumption that the near-IR light distribution of all ULIRGs near the beginning of the track is completely dominated by supergiants. That is generally not the case (Goldader et al. 1995). The model also requires that the (initially ~5 mag) brighter surface brightness of the young population (as compared to the old population) is exactly cancelled by obscuration in order to place the ULIRGs on the fundamental plane. The principal reason, however, is that ULIRGs are well separated from giant ellipticals in the third fundamental plane coordinate, $\sigma$, which should be little affected by differential extinction and population effects (Genzel et al. 2001). In particular, there are no conclusive theoretical or observational indications that the nuclear regions of young mergers are dynamically much colder than the regions further out in the systems. The simulations of Bendo & Barnes (2000) and Mihos (1999) indicate that the remnants are close to steady state by the time the two precursor nuclei have merged to <1 kpc separation. Velocity dispersions have a maximum at the center of the remnant and typically drop to <50% of the central value at $r \geq r_{eff}$, rather than the other way around.

Obviously, measurements of velocity dispersions in optically/UV selected QSOs are crucial to give an unambiguous answer to this question. No such data are available for QSOs at present, but Bettoni et al. (2001, and references therein) have compiled velocity dispersions for 73 radio galaxies in the redshift range 0.02 to 0.2. The mean (median) velocity dispersion of the radio galaxy sample is 256 (249) km/s, with a dispersion (uncertainty of mean) of 61 (7) km/s. These values are very close to those of local giant ellipticals ($M_B < -20.5$), which have a mean (median) dispersion of 269 (285) km/s, with a comparable scatter (Genzel et al. 2001). For comparison the mean (median) velocity dispersion of our ULIRG sample is 186 (165) km/s, with a dispersion (uncertainty of mean) of 55 (13) km/s, clearly much smaller. Fig.6 shows that even for the extreme assumptions of the toy model, ULIRGs do not evolve toward the location of the Bettoni et al. (2001) radio galaxies and of giant ellipticals if the currently measured velocity dispersion is constant, or does not drastically increase with time. We conclude that some motion of typical ULIRGs toward the upper right in Figure 5 is possible and indeed plausible, but that the average ULIRG observed in the local Universe does not evolve into a classical, optically bright QSO or radio galaxy.

## 4.3 The environments of ULIRGs, massive elliptical galaxies and optically bright QSOs

If ULIRGs, QSOs and elliptical galaxies trace a tight evolutionary sequence, one would also expect that they would all be found in similar types of environments. We have therefore investigated the environments in which local (z < 0.2) samples of ULIRGs, elliptical galaxies and quasars live. For this purpose we have used the 117 ULIRGs from the 1 Jy catalog of Kim & Sanders (1998) and the elliptical galaxy samples of Bender et al. (1992) and Faber et al. (1997). We have correlated the positions of the galaxies with the catalogs of galaxy clusters and groups available in NED to determine whether a given

ULIRG or elliptical lies in a rich cluster, a small group, or the field. For a coarse correlation we flagged any cluster or group located within 5 degrees on the sky and 0.02 in redshift of the ULIRG or elliptical. We then identified the galaxy as a cluster member if it was located within the cluster radius (as defined in the Abell catalog) and had a systemic velocity within the FWHM velocity dispersion of the cluster. For a galaxy located near a group we checked to make sure the galaxy was listed as a member of that group (usually from the lists of Huchra & Geller 1982 and Geller & Huchra 1983). McLure & Dunlop (2001) have done a complete clustering analysis on a sample of 44 quasars and radio galaxies at z~0.2 to study the cluster environments of radio-quiet and radio-loud AGN. They have determined spatial clustering amplitudes for their sample from deep HST imaging of the regions. This work is the most complete and careful analysis of host galaxy environments of local AGN available in the literature. For this reason we decided to use that study for comparison with the ULIRG and elliptical galaxy sample environments. It is not possible to make a full quantitative clustering analysis for the ULIRGs and elliptical galaxies, since to our knowledge no similarly complete imaging database is available for these systems.

The results of the environment comparison are shown in Figure 7, where we have placed each galaxy into one of three "richness" bins: (1) field galaxy or member of a small (<10 galaxies) group, (2) member of a cluster with Abell richness class 0, or (3) member of a cluster with Abell class >1.[4] We decided not to distinguish between field and small group membership because for the ULIRG sample we do not have access to wide-field deep imaging or lists of galaxy groups at suitably high redshifts.

---

[4] According to the definition of Abell (1958) a cluster has richness class 0 if $N_R > 30$ and a richness class $\geq 1$ if $N_R > 50$. Here $N_R$ is the richness count, and is the number of galaxies brighter $m_3+2$mag ($m_3$ is the third brightest cluster member) contained within a projected radius of $1.5h^{-1}$ Mpc of the cluster center.

Figure 7 shows that while elliptical galaxies and quasars are found in all environments, none of the 117 galaxies in the 1 Jy catalog of ULIRGs is located within a galaxy cluster. The environment distributions of the nearby giant elliptical galaxies and the McLure & Dunlop (2002) quasars are similar to each other. These sources are found in rich clusters as well as in the field and in small groups. The intermediate-mass, disky ellipticals, however, have an environment distribution that is more similar to that of the ULIRGs: they are preferentially found in the field.

The results of the environment comparison provide another piece of evidence that links the ULIRGs and intermediate-mass disky elliptical galaxies in an evolutionary sequence: both types of sources are preferentially found in the field (or in groups), and both avoid dense cluster environments. Since giant ellipticals and QSOs inhabit the field as well as rich cluster environments, we cannot rule out a link with ULIRGs based on environment alone. However, this analysis does tend to support the conclusions that we reached in the previous sections that there is no obvious direct evolutionary link from galaxies undergoing a ULIRG phase at the present epoch to the most massive elliptical galaxies, or optically bright QSOs. One cautionary note is that the redshift distributions of the ULIRG, elliptical galaxy and QSO samples are not identical. The mean redshifts of the McLure & Dunlop QSO sample and of the 1Jy ULIRG sample (Kim & Sanders 1998) are comparable (~0.2 and 0.15, respectively), but the elliptical galaxies studied by Bender et al. (1992) and Faber et al. (1997) mostly lie at z<0.03. Future work on well-studied samples of ellipticals, ULIRGs and AGN matched in redshift will provide more definitive answers to the question of whether the environments of these systems are systematically different.

### 4.4 ULIRG hosts, their central black holes and their evolution

Assuming now that the derived effective radii and velocity dispersions of the ULIRGs do in fact represent fair estimates of the host parameters, Table 4 summarizes the derived luminosities and masses of the ULIRG hosts, as well as the masses of their putative central black holes. The derived total absolute K magnitudes include corrections for an average effective K-band screen extinction of 0.7 mags. We obtained this correction factor from the '2kpc'-near-IR colors of Scoville et al. (2000, their Fig.7), in the limit of 'mixed' extinction and assuming the intrinsic colors for a stellar population of $10^7$ to $10^{8.5}$ years (see Genzel et al. 1998 for a discussion). A 2 kpc diameter aperture matches the average effective radius of the ULIRGs in Table 4, and the '2kpc'-colors thus should be a fair representation of the average near-IR light within $r_{eff}$. The integrated colors are quite similar to the 2kpc-colors for most ULIRGs. A screen extinction model cannot account for the observed extinction variations seen over a wide range of wavelengths in ULIRGs, while a mixed model fits most data reasonably well (Genzel et al. 1998), including the near-IR colors of Scoville et al. (2000). The average 2kpc-colors of the Scoville et al. sample correspond to $A_V$(mixed)~15±5, with Arp220 and IRAS17208-0014 representing the most extreme cases with $A_V$(mixed)≥50. The equivalent K-band screen extinction corresponding to $A_V$(mixed)=15 is $A_K$(eff)=0.7 (K-band extinction correction 1.9). We have applied this value to all galaxies of our sample, recognizing that it is probably conservative and represents an underestimate of the true extinction correction for several galaxies in Table 4. We have also corrected the absolute magnitudes in Table 4 for the contribution of an AGN, in cases where such a component was clearly present in the high resolution imaging data (Scoville et al. 2000, Lai et al. 1998). The absolute magnitudes listed in Table 4 should thus be fair representations of the stellar hosts. Our mean extinction corrections are greater than those of Colina et al. (2001) who concluded on the basis of optical extinction tracers that the near-IR extinction averaged over the hosts is negligible ($A_K(eff)<0.2$). In column 4 we list these extinction corrected stellar luminosities in units of the K-band luminosity at the knee of the local Schechter distribution function of early type galaxies ($M_K^*=-24.3$, after converting the values given by Loveday (2000) and Kochanek et al. (2001) to our cosmology). Column 5 lists the dynamical masses of the ULIRG hosts, obtained from our measurements of $\sigma$, $v_{rot}$ and $r_{eff}$ with the relationship of Bender et al. (1992):

$$m_{10} = c_2 (\sigma_{100})^2 r_{eff} \qquad (1),$$

where $m_{10}$ is the dynamical mass in units of $10^{10}$ $M_\odot$, $c_2$ is a structural constant, $\sigma_{100}$ is the velocity dispersion in units of 100 km/s, and $r_{eff}$ is the effective radius in units of kpc. For the ULIRGs we adopted King models with a ratio of tidal radius to core radius of about 50, corresponding to $c_2 \sim 1.4$ (Bender et al. 1992). In order to account for the contribution of rotation to the circular velocity, we used

$$\sigma^2 = \sigma^2_{obs} + a\, v^2_{obs,rot} \qquad (2),$$

where we took $a \sim 0.5$ for an inclination of $\sim 50^o$. In Table 4 we list these dynamical masses in units of the mass of an early type galaxy at the knee of the Schechter distribution, $m^* \sim 1.4 \times 10^{11} M_\odot$ (Bell and de Jong 2001; Cole et al. 2001). Column 6 then lists the K-band L/m ratio of our galaxies, in units of a early-type (old) L* galaxy. In column 7 we use the good correlation between nuclear black hole masses and galactic, stellar velocity dispersions in local early type galaxies and spheroids, as determined by Ferrarese and Merritt (2000) and Gebhardt et al. (2000), to estimate the central black hole masses in the ULIRG hosts. We adopt the relationship

$$M_{BH}(M_\odot) = 6.6 \times 10^6 \sigma_{100}^{4.3} \qquad (3).$$

Equation 3 is an average of the relationships given by Ferrarese and Merritt and by Gebhardt et al. The difference in the relationships derived by the two groups is largely due to the fact that they have extracted velocity dispersions over different sized apertures (see e.g. Tremaine et al. 2002). Ferrarese and Merritt derive velocity dispersions by extrapolating to $r_{eff}/8$, while Gebhardt et al. measure their velocity dispersions in apertures of $2\, r_{eff}$. We use the average value of the two relationships because we have extracted the velocity dispersions over apertures of diameters of $0.5$-$1.2 r_{eff}$, which are intermediate between the values used by the two groups.

Column 8 of Table 4 lists the efficiency of radiation of these putative black holes in units of the Eddington rate ($L/L_\odot = 3.3 \times 10^4 M_{BH}/M_\odot$), assuming (conservatively) that on average about 50% of the total (infrared) luminosity comes from accretion. This efficiency estimate is a lower limit for those ULIRGs with an energetically important AGN. In ULIRGs powered predominantly by star formation this fraction is less than 50% (Genzel et al. 1998), and thus the efficiency in column 8 is an upper limit. The case of Mkn 231 deserves special mention. The millimeter CO emission in this source reveals a rotating disk of very low ($\leq 20^\circ$) inclination (Downes and Solomon 1998). It is likely that the stars also are in this rotating disk, thus explaining the very low velocity dispersion. The true dynamical mass of the host of Mkn 231 may thus be larger than the value given in column 5 (by a factor of ~9: Appendix A), and the L/m, $m_{BH}$ and efficiency values would change to the values in parentheses in Table 4.

### 4.4.1. ULIRGs have 4-5 $L^*(K)_o$ hosts

Once dereddened, the ULIRG hosts are bright in the near-IR, with a median/mean (dereddened) K-band luminosity of about 4-5L*. Their median/mean K-magnitude is –25.8 with a dispersion of 0.8. The ULIRGs thus are comparable in their near-IR host properties to the radio-quiet QSOs, and somewhat less luminous than the radio-loud QSOs and radio galaxies in the nearby samples of Dunlop et al. (2002). Dunlop et al. (2002) find for their radio-quiet QSOs a median/mean K-magnitude of –26.0/-26.1, with a dispersion of 0.6; for their radio-loud QSOs a median/mean K-magnitude of –26.6/-26.5, with a dispersion of 0.5; and for their radio galaxies a median/mean K-magnitude of –26.2/-26.3, with a dispersion of 0.8. Our findings are also consistent with results from the much larger ULIRG sample of Sanders et al. (2000), who find from an analysis of 117 ULIRGs in the IRAS 1 Jy catalog that their mean observed (not extinction corrected) K-magnitude is about 2.7 $L^*(K)$. Correction for $A_K(eff)=0.7$ would bring that value to 5.1 $L^*(K)$, in excellent agreement with ours. The Dunlop et al. results in turn are in agreement with those for other local QSO samples (McLeod and Rieke 1994, 1995; Bahcall et al. 1997; Hutchings & Neff 1997; cf. Sanders et al. 2000). Our result appears at first inconsistent with the NICMOS H-band data of 27 ULIRGs from Colina et al. (2001) who find a mean absolute magnitude of $M_H$=-24.3, or 1.1$L_H$*. However, after converting to our definition of an L* galaxy ($M_H$*=-24.1), and correcting for $A_H(eff)=1$ (corresponding to $A_V(mixed)=15$), this value becomes 3.3 $L_H*$. While still smaller than our and the Sanders et al. (2000) means/medians, the two values are probably still consistent given the large sample variances and modest sample sizes.

**4.4.2. ULIRGs have ≤m* hosts**

The dynamical masses of ULIRG hosts are near to but somewhat lower than that of an m* elliptical host. The median/mean value of our sample is about $1.2 \times 10^{11}$ $M_\odot$, or about 0.85 m*. These dynamical masses also contain a significant contribution from interstellar molecular gas. Using the conversion from CO J=1-0 millimeter line flux to $H_2$(+He) mass proposed by Solomon et al. (1997) and Downes & Solomon (1998) for ULIRGs, we find that the average interstellar gas contribution to the dynamical mass (for those ULIRG sample sources with CO measurements) is about 10%, but can be as large as 30%-40% in a few cases (e.g. Mkn 231 and Arp220; Downes & Solomon 1998; Solomon et al. 1997). Accounting for gas, the average stellar mass of our ULIRG merger remnants is ~0.77m*. This result quantifies our earlier conclusion that ULIRG hosts have masses comparable to those of moderate velocity dispersion ellipticals/lenticulars, but are significantly less massive than giant ellipticals (5-15 m*). Combining the information from luminosity and mass, it is thus clear that the near-IR luminosity to stellar mass ratio of ULIRG hosts is on average 7 times (2 mag) greater than that of an old (>$10^{10}$ yr), early type galaxy. Applying Bruzual and Charlot (1993) evolutionary models, such an $L(K_o)/m(stars)$ ratio is typical of a stellar population of age ~few x $10^8$ years. This age is in rough agreement with the overall dynamical age of the merger remnants, but is somewhat larger than the current age of the active starbursts powering the far-IR luminosity (a few x $10^7$-$10^8$ years; Genzel et al.

1998).  This suggests that there may have been several star formation episodes during the merging process (Genzel and Cesarsky 2000; Rigopoulou et al 2000; Murphy et al. 2001), and that the near-IR light comes from a mixture of young and older stellar components. These considerations also quantify our earlier conclusion that the remarkable closeness of the young ULIRGs to the fundamental plane must at least in part be a cosmic conspiracy of extinction and population effects.

**4.4.3. ULIRGs radiate at ≥50% Eddington rate**

The black holes in ULIRG hosts with energetically dominant AGN are more efficiently fed than the black holes of local QSOs of the same luminosity.  Although there is a wide scatter observed in the accretion luminosity as a function of the velocity dispersion in both ULIRGs and QSOs, given the black hole masses estimated from the local black hole mass to central velocity dispersion relationship (Ferrarese and Merritt 2000; Gebhardt et al. 2000; Tremaine et al. 2002), the inferred central black holes in most ULIRGs would have to radiate at ~50% Eddington efficiency to account for as much as 50% of the far-IR luminosity (column 8 of Table 4).   The required efficiency is obviously less if we assume for a given ULIRG that recent star formation accounts for >50% of the IR luminosity. Typical efficiencies of the same nuclear luminosity range between 5 - 20% of the Eddington rate (Kaspi et al. 2000; Dunlop et al. 2002), albeit with a large scatter.  To be comparable with the radiation efficiencies of these local optically bright QSOs, the AGN in any of our sample ULIRGs could not contribute more than 10% of the total IR luminosity of these systems.  This is not likely to be the case, at least for the sources with energetically dominant AGN.  Many QSOs are located in gas poor, early-type hosts (Dunlop et al. 2002;

McLeod and Rieke 1995; Bahcall et al. 1997), although hosts of QSOs do span a wide range of structural characteristics. It is entirely plausible that QSOs in gas-poor hosts are only (re-) ignited to the same accretion luminosity as gas-rich ULIRGs if they have much bigger holes.

**4.4.4. ULIRGs cannot evolve into optically bright QSOs**

In summary, our main conclusion is that ULIRGs have less massive hosts, and thus less massive black holes than most optically bright, low-z QSOs/radio galaxies, in agreement with Colina et al. (2001). ULIRGs live in lower density environments than low-z QSOs/radio galaxies. Their black holes are more akin to those hosted by Seyfert galaxies than by QSOs. They can attain QSO-like bolometric luminosities because they accrete more and radiate more efficiently than average low-z QSOs. Once the enormous merger-induced influx of matter into the central black holes ceases, ULIRGs thus will not evolve into classical QSOs. They also cannot evolve into Seyfert nuclei, since the latter reside mostly in spiral hosts (McLeod and Rieke 1994, 1995). Local Universe ULIRGs may well evolve into inactive, moderately massive, field ellipticals or (if their black holes are fed) into objects similar to the hard-X ray luminous early type galaxies, which the Chandra observatory has recently discovered (e.g. Barger et al. 2001).

# 5. CONCLUSIONS

From our study of the kinematic properties of 18 ULIRG mergers with and without QSO luminosity AGN we find the following:

- AGN dominated ULIRGs have dynamical properties similar to those powered predominantly by star formation. Both types of late-stage ULIRG mergers are located very close to or in the fundamental plane of early type galaxies. Their properties are similar to those of moderate mass (~$L_*$) ellipticals/lenticulars with moderately fast rotation, but are significantly different from those of giant ellipticals. Giant ellipticals have larger velocity dispersions and effective radii than the ULIRGs.

- The bolometric and dereddened near-IR luminosities of our ULIRG sample are basically identical to those of the sample of 33 radio quiet/radio loud QSOs and radio galaxies studied by Dunlop et al. (2002), which are representative of other low-z optically selected QSO/radio galaxy samples. Yet ULIRGs and QSOs/radio galaxies are located in different parts of the $\mu_{eff}$-$r_{eff}$ projection of the fundamental plane. QSOs and radio galaxies are found near the locus of giant ellipticals, with typical effective radii of ≥10 kpc. In contrast, ULIRGs have effective radii of a few kpc or less, and much higher surface brightnesses. It is possible that some of the size and surface brightness differences between ULIRGs and optically selected QSOs/radio galaxies arise because ULIRGs contain compact, bright, circum-nuclear star forming regions that dominate the near-IR light at present but will fade away in a few Gyr. However, the large difference between the (large) velocity dispersions of giant ellipticals and radio galaxies, and the (modest) velocity dispersions of the ULIRGs should remain, since velocity dispersions should be much less affected by such evolutionary effects. We thus conclude that the hosts of present epoch ULIRGs and of QSOs/radio galaxies are very probably drawn from different populations. ULIRGs live in young, large $L_{IR}/m$, ~1m* hosts, while optically bright QSOs and radio galaxies on average live in old, small $L_{IR}/m$, ~10m* hosts, like giant ellipticals.

- ULIRGs and optically bright QSOs/radio galaxies also favor different density environments. ULIRGs and intermediate mass, disky elliptical galaxies preferentially reside in the field, with ULIRGs completely avoiding dense cluster environments. Giant elliptical galaxies and QSOs, on the other hand, are found in both dense cluster and field environments.

- From the host masses/velocity dispersions and the black hole mass to host mass/velocity dispersion ratios in local early type galaxies and bulges, we deduce that ULIRGs probably contain black holes of typical masses $\leq 10^8$ $M_\odot$, more akin to those in Seyfert galaxies than to those in QSOs. To attain their QSO-like far-IR luminosities ULIRGs (if powered by black hole accretion, such as those with energetically dominant AGN) must radiate at ~50% of the Eddington rate, significantly more efficiently than QSOs of comparable luminosities. Once the enormous merger induced influx of gas and dust into the central black holes subsides, ULIRGs may become inactive, moderately massive field ellipticals. In active periods when the black holes are fed, such galaxies may resemble the hard-X ray luminous, early type galaxies discovered recently by the Chandra observatory.

*Acknowledgments*: We thank the staffs of the Keck Observatory and the ESO VLT for their help with the observations. We are grateful to D. Sprayberry for providing us with the NIRSPEC saturation curve and to L. Tacconi-Garman for implementing that correction in an IRAF script. We thank N. Scoville for making available to us an electronic version of the NICMOS surface brightness distributions of several of our ULIRGs, and H.Dannerbauer for making available to us the near-IR images of IRAS14378-3651. We also appreciate very useful comments from the referee. This research has made use of the NASA/IPAC Extragalactic Database (NED), which is operated by the Jet Propulsion Laboratory, California Institute of Technology, under contract with the National Aeronautics and Space Administration.

# APPENDIX A: NOTES ON INDIVIDUAL SOURCES

IRAS 07598+6508 – This galaxy is a radio-quiet QSO with high and low ionization broad absorption lines (e.g. Lippari 1994; Hines & Wills 1999). Optical HST images reveal 2 clumps of emission located ~7" south and southeast of the QSO. These clumps may be due to emission from OB associations, and are evidence for very recent star formation (Canalizo & Stockton 2000). The NICMOS images of Scoville et al. (2000) are dominated by the bright point-like nuclear emission in all three IR bands, but there is also some fainter extended emission as well. This extended component could well be associated with the recent star formation studied by Canalizo & Stockton (2000). Deep R-band images show the presence of only one tidal tail, which has led Canalizo & Stockton to suggest that this ULIRG may be the result of the merger of a spiral and an elliptical galaxy. The high $z=0.149$ redshift of this source meant that we could no longer access the strong CO absorption bands in the H- or K-bands. We thus attempted stellar spectroscopy of weaker features in the J-band, but unfortunately, detected none.

IRAS 12071-0444 – Near-infrared adaptive optics imaging (Surace & Sanders 1999) of this system shows a bright compact nucleus together with more faint, extended emission from the host galaxy. Surace & Sanders find that the nuclear luminosity of this AGN ULIRG is similar to that of a moderately luminous Seyfert galaxy, rather than to that of a bona-fide QSO. Our J-band stellar velocity dispersion shows that this galaxy is dynamically very typical of the late-stage ULIRG mergers studied thus far. The limited signal-to-noise of the spectrum has not enabled us to measure a stellar rotational component.

Mkn 231-This source has the highest infrared luminosity ($L_{IR} \sim 3.2 \times 10^{12}$ $L_\odot$) of the original BGS ULIRG sample of Sanders et al. (1988). It also has the lowest velocity dispersion and most compact host of any ULIRG that we have studied thus far. As a result, Mkn 231 is an outlier to the lower left of the fundamental plane (Figs.3 to 5). The optical and near-IR images of this galaxy are dominated by the bright, point-like QSO nucleus (e.g. Lai et al. 1998; Surace & Sanders 1999; Canalizo & Stockton 2000), but also reveal prominent tidal tails extending out for over 75 kpc (e.g. Sanders et al. 1987). The central region also contains evidence for a significant amount of extended recent star formation (Krabbe et al. 1997, Taylor et al. 1999, Canalizo & Stockton 2000). Downes & Solomon (1998) have studied the molecular gas distribution and kinematics through sub-arcesecond resolution imaging of the millimeter CO emission in Mkn 231. This emission is distributed in a face-on (inclination $\leq 20^o$) rotating disk. It is possible the stars are also located within this rotating disk, thus explaining the very low velocity dispersion of the stars (115 ±10 km/s) that we derive from our NIR spectroscopy. If so, both the millimeter CO kinematics and our near-IR velocity dispersion indicate a dynamical mass of $\sim 1.3 \times 10^{11}$ $(0.34/\sin i)^2$ $M_\odot$ within the central R~1.5 kpc. Within this same region the K-band luminosity is $L_K \sim 1.6 \times 10^{10}$ $L_\odot$, implying a light-to-mass ratio of $\sim 0.15 (\sin i/0.34)^2$. Starburst models show that such a ratio will arise in a starburst that commenced $\sim 5 \times 10^8$ years ago (e.g. Tecza et al. 2000). For such a burst the models predict $L_{bol}/L_K \sim 70$. In the case of Mkn 231, this implies that the bolometric luminosity due to the host galaxy starburst is $\sim 1.1 \times 10^{12}$ $L_\odot$. If the correction factor between intrinsic circular velocity and observed dispersion is smaller than assumed above (e.g. for a more relaxed system or larger inclination), the light-to-mass ratio and fraction of the bolometric luminosity due to star formation will increase. The nonthermal radio flux density of the host and the FIR-radio correlation for starburst galaxies imply a star formation powered luminosity of $1.3 \times 10^{12} L_\odot$ (Lisenfeld, Völk & Xu 1996; Taylor et al. 1999). These results imply that the host galaxy contributes at least 1/3 of the total luminosity of this system.

The host structure and dynamics of Mkn 231 present an interesting puzzle for models of galaxy mergers. On the one hand, it has a low central stellar velocity dispersion and a very regularly rotating gas/stellar disk resembling those found in typical quiescent, intermediate-to-late type spiral galaxies rather than in merger remnants. On the other hand, there are very long and symmetric tidal tails seen in deep optical/NIR images indicating that this system is indeed in the late stages of a violent merger. Models predict the presence of such tidal tails in mergers of roughly equal mass galaxies (e.g. Barnes & Hernquist 1992; Mihos 1999). If that were the case here, the stellar velocity dispersion should be much higher than 115 km/s, comparable to the dispersions that we have measured in the other ULIRG mergers. Detailed models for Mkn 231 are needed to explain the peculiar dynamics of this system.

Mkn 273 – This well-studied merger contains a double nucleus, with a projected separation of 1" (0.7 kpc). NICMOS images show pronounced tidal tails in both the 1.1 micron emission map and the 2.2/1.1 micron obscuration image (Scoville et al. 2000). Much of the activity in the system is associated with the northern nucleus, where there is a considerable amount of young star formation. This is seen as extended emission associated with the northern nucleus in the NICMOS maps (Scoville et al. 2000), and as a group of compact radio sources in the VLBA map of Carilli & Taylor (2000). The northern nucleus is also the site of a massive gas disk seen in the high resolution HI absorption and radio continuum data of Carilli & Taylor (2000), as well as in the interferometric CO maps of Downes & Solomon (1998). The northern nucleus has also now been unambiguously identified as a Seyfert 2 nucleus in the hard X-ray Chandra observations of Xia et al. (2002).

The stellar velocity dispersion of the northern nucleus of Mkn 273 is among the highest seen in ULIRG mergers observed thus far. Only the double-nucleus merger NGC 6240, which has a massive self-gravitating gas concentration between its two massive nuclei, has a comparably large velocity dispersion (Tacconi et al. 1999; Tecza et al 2000). In the case of Mkn 273, however, we only see this large velocity dispersion in the northern nucleus, indicating that this system could be the result of an unequal mass merger in which one of the progenitors is a very massive, gas rich galaxy. We observe moderate rotation of both the stars and gas in the northern nucleus. The maximum rotation velocity is seen in our roughly east-west slit, consistent with dynamics derived from CO millimeter interferometry (Downes & Solomon 1998). However, Downes & Solomon find a peak rotation velocity of 280 km/s as compared with our derived rotation velocity of ~100 km/s for both the stars and the gas. The difference is likely due to a combination of 1) decoupling of the stellar and gas kinematics; 2) non-circular motions such as anisotropic outflow in a narrow-line region affecting the near-IR Fe line we use to measure gas rotation; and 3) the millimeter CO line emission arising primarily in a larger scale disk.

IRAS 14378-3651: HST I- and H-band images of this galaxy reveal a single nucleus, elliptical-like galaxy with multiple shells and spiral arms (Bushouse et al. 2002). This system is clearly in a very advanced state of merging.

IRAS 15250+3609: The NICMOS image of this system reveals a dominant nucleus and a fainter possible second nucleus located 0.7" to the southeast. The rotational component of 60 km/s that we observe is along the line joining the 2 NICMOS components.

Figure 1. H- and K-band surface brightness distributions for 5 of the new program galaxies, along with the best-fitting de Vaucouleurs, $r^{1/4}$-profiles (black, dashed). The derived effective radii are listed in each inset. The data are from Scoville et al. (2000, NICMOS: IRAS 07598+6508, Mkn 273, IRAS 15250+3609); from Dannerbauer (private communication, NTT SOFI: IRAS 14378-3651); and from Lai et al. (1998, CFHT PUEO: Mkn 231).

Figure 2. NIRSPEC and ISAAC H- and J-band spectra of six program galaxies (continuous curves), plotted with K/M stellar template spectra convolved with the Gaussian best reproducing the line-of-sight-velocity-distribution obtained from our Fourier-quotient analysis (dashed curves). Spectra are typically extracted from a 1" aperture centered on the nucleus; for Mkn 231, we have taken an annulus with an inner radius of ~0.3" and an outer radius of ~0.7".

Figure 3. Projections of the fundamental plane. Red squares (and error bars) are ULIRGs, combining the present 7 sources with 11 sources from Genzel et al. (2001). ULIRGs with strong or energetically dominant AGN are marked with 'A'. Grey crosses are ellipticals from the compilation of Pahre (1999); green filled circles mark the LIRG mergers from Shier & Fischer (1998) and James et al. (1999). Light blue squares with crosses are giant boxy ellipticals (M(B)<-20.5, B-K=3.9), and dark blue filled circles mark disky ellipticals (M(B)<-18, B-K=3.9) both from Bender et al. (1992) and Faber et al. (1997). Left: edge-on view of the fundamental plane. The dashed line is the best-fit slope of the fundamental plane from the K-band studies of Mobasher et al. (1999) and Pahre (1999): log $r_{eff}$= 1.4

(log σ + 0.22 μ$_{eff}$ (K)) + const. Also indicated are the effects of extinction (bottom) and age/evolution of the stellar population (top). Top right: r$_{eff}$-σ projection of the fundamental plane. Bottom right: σ-r$_{eff}$-v$_{rot}$/σ distribution of the galaxies.

Figure 4. Comparison of the ULIRG mergers with optically bright QSOs. Left: μ$_{eff}$(K)-r$_{eff}$ distribution. AGN from the Dunlop et al. (2002) local sample (z<0.3) are plotted with light blue symbols: 'Q' marks radio-quiet QSOs, 'L' marks radio loud QSOs and 'G' marks radio galaxies. The additional symbols are as for Figure 3. For the ULIRGs, the bolometric luminosity is a sum of the 1-1000μm infrared luminosity and (a small) contribution from the optical band. For the Dunlop et al. QSOs we estimated L$_{bol}$ from nuclear R-band luminosities and a bolometric correction of 2.9 mag, as given in Elvis et al. (1994). The ULIRGs and QSOs have nearly identical luminosity distributions, consistent with the Sanders et al. (1988) scenario.

Figure 5. Evolution of merger remnants in the log r$_{eff}$-μ$_{eff}$ plane. Symbols and letters are the same as in Figures 3 and 4. In addition, the green curve shows the evolution of a toy model merger remnant consisting of two $r^{1/4}$-components: an old component with r$_{eff}$=10 kpc, and a young component with r$_{eff}$=0.4 kpc. For a young merger remnant with an initial μ$_{eff}$=14 (the location of several of the ULIRG mergers), the curve (with labels indicating log t (yrs)) shows the evolution of the composite merger as the young component fades according to Bruzual & Charlot (1993) models.

Figure 6. Evolution of merger remnants within the fundamental plane from a viewing position perpendicular to the plane: Symbols and letters are the same as in Figures 3, 4 and 5. The dark brown curve (plus triangles) shows the evolution of the toy model merger remnant displayed and described in Figure 5. The evolution of the composite merger assumes that the velocity dispersion remains constant ($\sigma$=185 km/s) with time. The filled green circles mark the locations of the radio galaxies from Bettoni et al. (2001, and references therein). It is clear that ULIRGs cannot evolve into giant ellipticals or radio galaxies, even under the extreme assumptions of the toy model (see text).

Figure 7. The environments of ULIRGs compared with those of elliptical galaxies and QSOs. The ULIRG sample is the entire 1 Jy catalog (118 galaxies) from Kim & Sanders (1998); the 35 giant, boxy ellipticals and 32 intermediate-mass, disky ellipticals are from Bender et al. (1992) and Faber et al. (1997); and the 44 radio-loud and radio-quiet quasars and radio galaxies are from McLure & Dunlop (2001). The three environment categories shown here are: 1) field or small group with less than 10 members; 2) cluster with Abell richness class 0; and 3) cluster with Abell richness class 1 or greater.

TABLE 1
Source List

| Source | R.A. (2000) | Dec. (2000) | Z | Scale (kpc/arcsec) | log ($L_{IR}$) | p.a. slits | Integr. time (hours) | Instrum. |
|---|---|---|---|---|---|---|---|---|
| 07598+6508 | $08^h04^m33.1^s$ | +64°59'49" | 0.149 | 2.49 | 12.4 | +20° | 1 | NIRSP |
| 12071-0444 | 12 09 45.1 | -05 01 14 | 0.129 | 2.22 | 12.3 | +20,-10 | 0.7,0.7 | NIRSP |
| 12540+5708 (Mkn 231) | 12 56 14.2 | +56 52 25 | 0.042 | 0.82 | 12.5 | 10,-30,-80 | 0.7, 0.7,0.7 | NIRSP |
| 13428+5608 (Mkn 273) | 13 44 42.1 | +55 53 13 | 0.037 | 0.73 | 12.1 | +15, +95 | 0.7, 0.7 | NIRSP |
| 14378-3651 | 14 40 58.9 | -37 04 33 | 0.068 | 1.27 | 12.1 | -45 | 1.3 | ISAAC |
| 15250+3609 | 15 26 59.4 | +35 58 38 | 0.055 | 1.05 | 12.0 | +45,-45 | 0.7,-0.7 | NIRSP |
| 19254-7245 | 19 31 21.6 | -72 39 25 | 0.061 | 1.16 | 12.0 | -13 | 0.5 | ISAAC |

TABLE 2
Dynamical/Structural Properties of ULIRG Mergers[a]

| Galaxy | z | $\sigma$ stars | $v_{rot}$ stars | $v_{rot}/\sigma$ stars | $v_{rot}$ gas | $v_{gas}/v_{stars}$ | $r_{eff}$ kpc | $\mu_{eff}(K)$ (mag) | $M(K)_{tot}$ (mag) | notes |
|---|---|---|---|---|---|---|---|---|---|---|
| I 07598+6508* | 0.149 | - | - | - | - | - | 22(5) | 18.5 | -27.0 | b |
| I 12071-0444* | 0.129 | 200(100) | - | - | - | - | - | - | - | c |
| Mkn 231* | 0.042 | 120(10) | 25(10) | 0.21 | 25(10) | 1.0 | 0.3(0.1) | 11.0 | -24.8 | d |
| Mkn 273* | 0.037 | 285(30) | 110(20) | 0.39 | 100(10) | 0.9 | 2.0(0.5) | 14.8 | -24.8 | e |
| I 14378-3651* | 0.068 | 153(10) | 15(10) | 0.10 | - | - | 0.7(0.1) | 13.6 | -24.6 | f |
| I 15250+3609 | 0.055 | 150(10) | 60(15) | 0.40 | 60(10) | 1.0 | 1.2(0.1) | 15.2 | -24.2 | g |
| I 19254-7245* | 0.061 | 188(10) | - | - | - | - | - | - | - | h |

a) An asterisk (*) behind the source name in the first column denotes that the source contains a strong AGN. The (central) stellar velocity dispersion (in the third column) is typically in an aperture of 0.6"x (1-1.5"). Rotation velocities of stars (fourth column) and emission line gas (sixth column) are half of the intensity-weighted velocity differences on either side of the galaxy, at radii of 1-2". They do not include corrections for inclination and beam smearing and represent lower limits to the true rotation speed. Numbers in parentheses are the uncertainties (±), including systematic effects. The uncertainty of the average surface brightness within the effective radius (eighth column: from fit to de Vaucouleurs $\{r^{1/4}\}$ law), $\mu_{eff}$ ($r \leq r_{eff}$) (ninth column: in K-band), is about 0.7 mag. $M(K)_{tot}$ (tenth column) is the absolute K-band magnitude of the entire merger system ($M(K)_* \sim -24.3$). In systems with bright AGN the nuclear near-IR emission was subtracted/discarded before host surface brightness, etc. were estimated.
b) NIRSPEC J-band data dominated by strong and broad 1.083μm HeI; no stellar absorption features detected. Surface brightness data from H- and K- NICMOS observations of Scoville et al. (2000).
c) From J-band stellar spectroscopy.
d) H-band data. Surface brightness data from the CFHT PUEO observations of Lai et al. (1998).
e) H-band data. Surface brightness data from H- and K- NICMOS observations of Scoville et al. (2000).
f) H-band data. Surface brightness data from NTT K-band observations (Dannerbauer, priv.comm.).
g) H-band data. Surface brightness data from H- and K- NICMOS observations of Scoville et al. (2000).
h) H-band data, south nucleus only.

# TABLE 3

Summary of Structural and Dynamic Parameters of ULIRG Sample[a]

| 1 | 2 | 3 | 4 | 5 | 6 | 7 |
|---|---|---|---|---|---|---|
| Galaxy | z | $\log\sigma$ (km/s) | $\log(r_{eff})$ (kpc) | $\log(v_{rot}/\sigma)$ | $\mu(K)_{eff}$ (mag/arcs$^2$) | $\log\sigma+0.22\mu(K)_{eff}$ |
| IRAS00262+4251 | 0.0970 | 2.23(0.04) | 0.53(0.13) | -1.1(0.4) | - | - |
| IRAS00456-2901 | 0.1098 | 2.21(0.07) | 0.11(0.17) | -0.56(0.1) | 15.8(0.7) | 5.69(0.16) |
| IRAS01388-4618 | 0.0912 | 2.16(0.03) | -0.02(0.18) | -0.04(0.06) | 13.7(0.7) | 5.17(0.14) |
| Mkn1014 | 0.1630 | 2.30(0.13) | 0.60(0.22) | - | 16.5(0.7) | 5.93(0.19) |
| IRAS07598+6508 | 0.1490 | - | 1.34(0.09) | - | 18.5(0.7) | - |
| Mkn231 | 0.0424 | 2.08(0.02) | -0.52(0.10) | -0.68(0.09) | 11.0(0.7) | 4.50(0.14) |
| Mkn273 | 0.0371 | 2.45(0.05) | 0.30(0.15) | -0.41(0.09) | 14.8(0.7) | 5.71(0.15) |
| IRAS14348-1447 | 0.0824 | 2.22(0.05) | 0.48(0.14) | -0.52(0.14) | 17.0(0.7) | 5.96(0.15) |
| IRAS14378-3651 | 0.0680 | 2.18(0.03) | -0.15(0.06) | -1.0(0.3) | 13.6(0.7) | 5.17(0.14) |
| IRAS15250+3609 | 0.0550 | 2.18(0.03) | 0.08(0.04) | -0.40(0.1) | 15.2(0.7) | 5.52(0.14) |
| Arp220 | 0.0178 | 2.22(0.03) | -0.26(0.24) | -0.04(0.12) | 13.9(0.7) | 5.28(0.14) |
| NGC6240 | 0.0248 | 2.51(0.04) | 0.04(0.12) | -0.18(0.09) | 13.7(0.7) | 5.52(0.15) |
| IRAS17208-0014 | 0.0428 | 2.36(0.03) | 0.15(0.09) | -0.32(0.08) | 14.6(0.7) | 5.57(0.14) |
| IRAS20087-0308 | 0.1057 | 2.34(0.03) | 0.52(0.13) | -0.64(0.13) | 15.9(0.7) | 5.84(0.14) |
| IRAS20551-4250 | 0.0431 | 2.15(0.05) | 0.32(0.14) | -0.54(0.12) | 16.4(0.7) | 5.75(0.15) |
| IRAS23365+3604 | 0.0645 | 2.16(0.04) | 0.68(0.09) | -1.0(0.44) | 17.0(0.7) | 5.90(0.15) |
| IRAS23578-5307 | 0.1250 | 2.28(0.16) | 0.96(0.24) | - | 18.1(0.7) | 6.26(0.21) |

(a) numbers in parentheses are 1σ errors

TABLE 4

Derived Properties of ULIRG Hosts

| 1 | 2 a | 3 b | 4 c | 5 | 6 d | 7 e | 8 | 9 f |
|---|---|---|---|---|---|---|---|---|
| Galaxy | $\log(L_{bol})$ | $M(K)_0$ | $L(K)_0/L^*$ | $m_{dyn}/m^*$ | $m_{gas}/m^*$ | $L(K)_0/m$ $(L^*/m^*)$ | $\log m_{BH}$ $(M_\odot)$ | Eddington Efficiency |
| IRAS00262+4251 | 12.1 | - | - | 1 | 0.07 | - | 7.8 | - |
| IRAS00456-2901 | 12.3 | -26.1 | 5.2 | 0.35 | - | 16 | 7.7 | 0.6 |
| IRAS01388-4618 | 12.1 | -26.5 | 7.6 | 0.3 | - | 30 | 7.5 | 0.6 |
| Mkn1014 | 12.6 | -26.6 | 8.3 | 1.6 | 0.06 | 5.5 | 8.1 | 0.5 |
| IRAS07598+6508 | 12.6 | -27.7 | 23 | - | - | - | - | - |
| Mkn231 (g) | 12.7 | -25.5 | 3.0 | 0.04(0.35) | 0.26 | 100(12) | 7.1(8.0) | 6 (0.7) |
| Mkn273 | 12.2 | -25.5 | 3.0 | 1.7 | 0.03 | 1.8 | 8.8 | 0.04 |
| IRAS14348-1447 | 12.4 | -25.8 | 4.0 | 0.8 | - | 5 | 7.8 | 0.7 |
| IRAS14378-3651 | 12.2 | -25.3 | 2.5 | 0.2 | - | 18 | 7.6 | 0.6 |
| IRAS15250+3609 | 12.1 | -24.9 | 1.7 | 0.3 | - | 7 | 7.6 | 0.5 |
| Arp220 | 12.2 | -24.6 | 1.3 | 0.2 | 0.39 | 10 | 7.8 | 0.4 |
| NGC6240 | 11.9 | -25.8 | 4.0 | 1.4 | 0.07 | 3 | 9.0 | 0.02 |
| IRAS17208-0014 | 12.4 | -25.7 | 3.6 | 0.8 | 0.10 | 5 | 8.4 | 0.2 |
| IRAS20087-0308 | 12.5 | -26.2 | 5.8 | 1.6 | 0.12 | 4 | 8.3 | 0.25 |
| IRAS20551-4250 | 12.1 | -25.2 | 2.3 | 0.4 | - | 6 | 7.4 | 0.7 |
| IRAS23365+3604 | 12.2 | -25.9 | 4.4 | 1.0 | 0.10 | 5 | 7.5 | 0.7 |
| IRAS23578-5307 | 12.2 | -26.2 | 5.8 | 3.3 | - | 2 | 8.0 | 0.2 |
| Mean | 12.3 | -25.8 | 5.3 | 0.9 | 0.13 | 9 | 7.9 | 0.8 |
| Median | 12.2 | -25.8 | 4.0 | 0.8 | 0.10 | 6 | 7.6 | 0.5 |
| Dispersion | 0.2 | 0.7 | 5.1 | 0.8 | 0.12 | 8 | 0.5 | 1.5 |

a) $A(K)_{eff}$=0.7 mag.
b) $M^*(K)$=-24.3
c) $m^* = 1.4 \times 10^{11} M_\odot$
d) $H_2$ (+He) gas masses from CO J=1-0 observations of Solomon et al. (1997) and Downes & Solomon (1998), and using a conversion factor of 1.4 $M_\odot$/[K km/s pc$^2$] from CO line flux to gas mass.
e) Corrected for gas mass in column 6; where no information on the gas mass was available we used the median value (0.1) of column 6.
f) for $L_{AGN} = 0.5 L_{bol}$
g) for Mkn 231 numbers in parenthesis assume sin i=0.34, and thus an 8.6 times greater dynamical mass than that given in Appendix A. The reduction of the Eddington efficiency to a value <1 for this inclination strengthens the case for adopting it.